\documentclass[runningheads]{cl2emult}

\usepackage{makeidx}  
\usepackage{graphicx} 
\usepackage{cropmark} 
\usepackage{eso}      
\makeindex            

\begin{document}
\title*{The DEEP2 Redshift Survey }
\toctitle{The DEEP2 Redshift Survey}
%
%
\titlerunning{The DEEP2 Redshift Survey}
%
\author{Marc Davis \& Jeffrey A. Newman\inst{1}
\and Sandra M. Faber \& Andrew C. Phillips\inst{2}}

\authorrunning{Davis {\it et al}}
%
%
\institute{University of California, Berkeley, CA 94720, USA
\and Lick Observatory, University of California, Santa Cruz, CA, USA}

\maketitle              

\begin{abstract} The DEIMOS spectrograph is nearing completion in the UC
Lick shops and will be delivered to the Keck Observatory in 2001. Once it
is operational, a team of astronomers will initiate DEEP2, a major
redshift
survey of galaxies that will consume approximately 120 nights at the
Keck Observatory over a
three year period.  Our goal is to gather high-quality spectra of $\approx
60,000$ galaxies with $Z>0.7$ in order to study the
properties
 and large scale clustering of galaxies at $Z \approx 1$.  The survey will
be done at high spectral resolution, $R=\lambda/\Delta \lambda \approx
3500$, allowing us to work between the bright OH sky emission lines and to
infer linewidths for many of the target galaxies.  The linewidth data will
facilitate the execution of the classical redshift-volume cosmological
test, which can provide a precision measurement of the equation of state
of the Universe.  \end{abstract}

\section{The DEIMOS spectrograph}
DEIMOS (the DEep Imaging Multi-Object Spectrograph) 
is intended for imaging and multislit spectroscopy
over a field of view that is approximately a rectangle of size 16' by 5'.
At the heart of DEIMOS is a massive 9-element refractive camera, with
three aspherical surfaces and the largest piece of CaF$_2$ ever cast
(13.5" diameter).  The focal plane of the camera 
is an 8k-by-8k pixel array of eight 2k-by-4k CCDs manufactured 
by the MIT Lincoln Laboratory,
with 15$\mu$ (0.119 arcsec) pixels.  The RMS image diameter 
of the spectrograph itself is $\sim 26\mu$, or 0.21 arcsec.  
We have in hand a full complement of thinned CCDs, and we expect to soon have
a full array of high resistivity $30\mu $thick CCDs, which have much 
enhanced QE beyond 9000 $\AA$ and less fringing than the thin chips.

 DEIMOS
is too massive for the Cassegrain focus and 
will reside on a Nasmyth deck of the Keck-II telescope, but since
the field rotates as the telescope tracks across the sky, DEIMOS  must 
rotate along its long axis.  Changing gravitational forces will 
inevitably cause any instrument to flex, but maintaining constant 
wavelength to pixel registration is essential for careful flat fielding
and sky subtraction.  DEIMOS has been designed with
an active 2-D
flexure compensation system  in which a tent mirror and the
X-coordinate of the detector package can be slightly adjusted  by means of
a closed-cycle feedback loop. The goal is for the  
wavelength--to--pixel 
registration within DEIMOS to be held to  within  0.1 pixel
RMS for long periods of time, hopefully obviating the need to take extensive
calibrations frequently.

DEIMOS is being built as a facility instrument for Keck and will have a
grating slide with room for three diffraction gratings plus a mirror
for imaging.  The slit masks are stored in a juke-box with room for
13 separate masks.  The masks are thin aluminum plates in which slitlets
will be cut by a computer controlled milling machine. These slits
can be tilted to follow the major axis of target galaxies, so as to 
facilitate measurement of rotation curves. The masks are stored
flat but upon insertion they 
are bent to a cylindrical shape that  closely approximates
 the spherical focal plane of Keck.

\section{The DEEP2 Observing Team}

Once DEIMOS is operational, we shall commence DEEP2, a survey of faint
galaxies designed to characterize galaxies and the galaxy distribution at
a redshift $Z=1$.  (DEEP1 is the project described by Koo at this meeting.)
The DEEP2 goal is to generate a sample of uniform
quality data with a well defined selection criterion that will be suitable
for many different analyses.  This is a collaborative project amongst
astronomers at UC, Caltech, and the Univ of Hawaii, in addition to outside
collaborators.  Team members with Keck access are M. Davis, S. Faber, D.
Koo, R. Guhathakurta, C. Steidel, R. Ellis, J. Luppino, and N. Kaiser.
This program will use 120 Keck nights over a three year period. 

\subsection{Fields and Photometry}
\begin{table}
\caption{Fields Selected for the DEEP2 Survey}
\begin{center}
\begin{tabular}{|c|c|c|c|} \hline
RA & dec & (epoch 2000) & mask pattern \\ \hline
14$^h$ 17 & +52$^\circ$ 30  &  Groth Survey Strip  & 120x1 \\ \hline
16$^h$ 52 & +34$^\circ$ 55   & last zone of low extinction & 60x2\\ \hline
23$^h$ 30 & +0$^\circ$ 00   & on deep SDSS strip & 60x2 \\ \hline
02$^h$ 30 & +0$^\circ$ 00   & on deep SDSS strip & 60x2 \\ \hline
\end{tabular}
\end{center}
\end{table}

The DEEP2 survey will be undertaken
in four fields, as listed in Table 1.  The fields were chosen as low
extinction zones that are continuously observable at favorable zenith 
angle from Hawaii over a six month interval. 
One field is the Groth
Survey strip, which has good HST imaging and which will be the target of very 
deep IR imaging by SIRTF, and two of the fields are on the
equatorial strip that will be deeply surveyed by the Sloan Digital Sky Survey
(SDSS) project \cite{york}.  Each of these fields is the target of a CFHT 
imaging survey
by Luppino and Kaiser, whose primary goal is very deep imaging for weak 
lensing studies.  They are using the new UH camera (8k by 12k pixels) with
a field of view of 30' by 40' in the B, R, and I bands.
The imaging will be obtained in random pointings spread over
a field of $3^\circ$ by $3^\circ$, but with continuous coverage of a strip
of length $2^\circ$ by $30'$ in the center of each field.

Given this enormous photometric database, we shall use the color information
to select galaxies with $Z>0.7$.  DEIMOS will then be used
to undertake a spectroscopic survey of galaxies with $m_I(AB) \le 23.5$ and
meeting our color criteria.  At this relatively bright flux limit, 60\% of
the
galaxies will have $Z<0.7$ \cite{lilly,cohen}; the photometric
redshift preselection eliminates this foreground subsample, 
allowing the DEEP2 project to focus its effort on the high-redshift
Universe.

\subsection{Choice of Grating and Spectral Resolution}
We anticipate that the
workhorse grating for the DEEP2 survey will be the 900 lines/mm grating, 
which has an anamorphic factor of 1.4.  This grating 
will provide spectral coverage of 3500 $\AA$ in one setting.  If we use
slits of width 0.75", they will project to a size of 4.6 pixels, or
a wavelength interval of 2 $\AA$.  Thus the resolving power of the observations
will be quite high, $R \equiv \lambda / \Delta\lambda =3700$.

The MIT-LL CCDs have exceptionally low readout noise, 1-2 $e^-$, and the time
to become sky-noise limited is less than 10 minutes, 
even when using the 1200 l/mm grating.  The large number
of pixels in the dispersion direction allows high resolution with
substantial spectral range,  so that we can work between  the bright OH
sky emission lines while remaining sky-noise limited.

We will set the grating tilt so that the region 6000-9000 $\AA$ is centered on
the detector, thus assuring that the 3727 $\AA$ [OII] doublet is in range
for galaxies with $0.7 < Z < 1.2$.  At the planned spectral resolution, 
the velocity resolution will be 80 km/s.  
The [OII] doublet will thus be resolved for all the galaxies,
giving confidence to the redshift determination even if no other features are 
observed.  With sufficient flux it should be possible to measure the velocity 
broadening of the lines, which will hopefully lead to an estimate of the
gravitational potential-well depth of a substantial fraction of the 
galaxies within the survey.

\subsection{Observing Strategy--Target Selection }

In each of the four selected fields of Table 1, we shall densely target a
region of 120' by 16' or 120' by 30' for DEIMOS spectroscopy. 
We intend to produce
120 separate masks per field; each mask 
will each contain slitlets selected from  
a region of size 16' by 4', 
with the slitlets mostly aligned 
along the long axis, but with others tilted as much as $30^\circ$ to track
extended galaxies.  Our goal is to select an average of 
 130 slitlets per mask, selected from the
those galaxies 
meeting our flux and color cuts.  The mean surface
density of candidate galaxies  exceeds the number of objects we
can select  by approximately 30\%, 
and spectra of selected targets cannot be allowed to overlap.  However, 
this will not cause problems with subsequent analyses if we take account
of the positions of those galaxies for which we did not obtain spectroscopy.
In the Groth strip region, because of the interests of other 
collaborative scientific
projects, our plan is to construct 120 distinct masks each
offset from their neighbor 
by 1', and to select targets without regard to color.  Thus, any spot
on the sky will be found within 4 masks, giving every galaxy
 4 chances to appear
on a mask without conflict. In the other three survey fields, we plan 
to use the color selection, thus halving the source density of targets,
and
to step 2' between masks, giving a galaxy two chances to be selected without
conflict.  In these fields the masks will form a pattern of 60 by 2, covering
a field of 120' by 30'.  At $Z \approx 1$, this field subtends a comoving
interval of $80~x~20 h^{-2}$ Mpc$^2$, and our redshift range translates to
a comoving interval of $\approx 800h^{-1}$ Mpc (flat Universe with 
$\Omega_\Lambda =0.7$).

Note that this survey differs from the planned VLT/VIRMOS survey 
\cite{lefevre} in that a DEIMOS spectrum will occupy a full row of the CCD
array (8k pixels), so the VLT/VIRMOS  multiplexing will be higher than ours. 
The DEEP2 spectra will have resolution 20 times higher than planned for the
bulk of the VLT/VIRMOS project, but our volume surveyed will be smaller
(partially because we are excluding objects with $Z<0.7$).


\subsection{Two Surveys in One}  
The DEEP2 project is actually subdivided into two surveys, 
the 1HS (one hour [integration time]
survey) and the 3HS (three hour survey).  The 1HS is the backbone of the 
DEEP2 project and will require 90 nights on Keck for execution.  The planned
large scale structure studies are the main drivers of the 1HS design.  The 
3HS will require 30 nights of Keck time to obtain spectra of $\approx 5000$ 
targets. The major
scientific focus of the 3HS is to study 
the properties of galaxies, but it will also serve a critical role of 
checking the quality of linewidths and other properties derived from 1HS 
spectra.  3HS targets will be selected to a
flux limit up to one magnitude fainter than the 1HS.
The 3HS  survey fields will be limited to a few 16' by 4'
regions, plus one  field 16' by 16' in the center of the Groth Survey zone.
We intend to acquire HST imaging in all 3HS fields, as well as Chandra
and XMM images.


\subsection{Keeping up with the Data}

The data rate from DEIMOS will be in excess of 1 Gbyte/hour, so 
automated  reduction and analysis tools are absolutely imperative.  We
have been working closely with the SDSS team and intend to adapt the
IDL code of Schlegel and Burles for our pipeline reduction.  
The photometric and
spectroscopic databases for the project are currently planned to be
IDL structures,
stored on disk as FITS binary tables.  Although the total raw data will exceed
one Tbyte, the reduced data will be modest in size by today's standards,
$< 50$ Gbytes.

\section{Science Goals of DEEP2}

The 1HS survey is designed to provide a fair sample volume for analysis
of LSS statistical behavior, particularly for clustering studies
on scales $ < 10h^{-1} $ Mpc. 
 The comoving volume surveyed in the 1HS program will
exceed that of the LCRS survey \cite{shectman}, a survey 
 which has proven to be an 
outstanding resource for low redshift studies of LSS. 
When the data is in hand, we plan a number of scientific analyses, with
major programs listed below.

\begin{itemize}
\item
Characterize the linewidths and spectral properties of galaxies versus color,
luminosity, redshift, and other observables. The 3HS is designed
to reach to a depth to allow more detailed analysis of the internal properties
of galaxies at high redshift, including rotation curves, linewidths from 
absorption spectra, and stellar populations.

\item 
Precisely measure the two--point and three--point correlation functions of 
galaxies at $Z=1$ as a function of other observables, 
such as color, luminosity,
or linewidth.  For the higher--order correlations,  dense sampling is
essential.  Observations of Lyman limit galaxies at $Z=3$ \cite{steidel}
show that the bias in the galaxy distribution was considerably higher in
the past.  Higher order correlations in the galaxy distribution are one way
to estimate the presence of bias in the galaxy distribution \cite{fry}. If the
galaxy bias is larger at $Z=1$ than at present, the correlation
strength of different subsamples of galaxies should show more systematic
variation than is observed for galaxies at $Z=0$.

\item
Measure redshift space distortions in the galaxy clustering at $Z=1$ by
means of the $\xi(r_p,\pi)$ diagram and by other  measures.  The
evolution of the thermal velocity dispersion is another handle that can
separate the evolution of the galaxy bias from the evolution of the 
underlying matter distribution.  The redshift precision afforded by
the resolution of DEIMOS will make this measurement possible.

\item 
Count galaxies as a function of redshift and linewidth, in order to 
execute the classical redshift-volume cosmological test.  Details are given by
\cite{newman} who show that this test can provide a precision measurement of
the cosmological parameters, $\Omega_m$ and $\Omega_\Lambda$. If instead 
one assumes the universe to be flat, the test can set a strong constraint
on the equation of state parameter $w$ of the dark energy component, in which
$P= w\rho$ \cite{steinhardt}

\item
Separate the Alcock-Paczynski effect \cite{alcock},\cite{ball} from 
the redshift 
space distortions of the $\xi(r_p,\pi)$ diagram.  This effect relates
intervals of angular separation versus intervals of redshift separation 
as a function of redshift. 
An object that appears spherical at low redshift would
appear elongated in redshift at $z>0$, but the degree of elongation is
a function of $q_0$.  This effect is 10 times smaller in amplitude than
the redshift-volume effect and will be challenging to measure, but 
it is conceivable that the DEEP2 project will
provide data that can measure this effect and separate it from the
other expected redshift space distortions.

\end{itemize} 
\section{Conclusions}
The flood of data now coming from the SDSS and 2DF 
projects detailing
the local Universe will soon be
complemented by data from the VLT/VIRMOS project and by the DEEP2 survey 
providing detailed information on the Universe at $Z \approx 1$,
thus continuing the revolution in precision cosmology and large-scale 
structure.  We intend to share our results with the public and to put our
spectra online in a timely manner.  Further details on the survey can be
found at the web site
http://astro.berkeley.edu/deep/.  The coming years promise 
to be extraordinarily busy for us, but very exciting.

\subsection{Acknowledgements}

{This work was supported in part by NSF grants AST00-71048 
and AST95-29028. The DEIMOS spectrograph is funded by a grant from CARA
(Keck Observatory), by an NSF Facilities and Infrastructure grant (AST92-2540), by the Center for Particle Astrophysics, and by gifts from Sun Microsystems
and gifts from Sun Microsystems and the Quantum Corporation.}

\clearpage
\addcontentsline{toc}{section}{Index}
\flushbottom
\printindex

\end{document}